\documentclass[aps,pra,twocolumn,groupedaddress]{revtex4-1}

\usepackage[cp1250]{inputenc}
\usepackage{amssymb}
\usepackage[dvips]{graphicx}
\usepackage[dvips,usenames]{color}
\usepackage{hyperref}
\usepackage{amsfonts}
\usepackage{nicefrac}
\usepackage{graphicx}
\usepackage{amsmath}
\usepackage{dsfont} 

\global\long\def\ket#1{\mbox{\ensuremath{\left|#1\right\rangle }}}

\global\long\def\braket#1#2{\left\langle#1\vphantom{#2}\right|\left.#2\vphantom{#1}\right\rangle }

\global\long\def\Braket#1#2#3{\left\langle #1\vphantom{\{#2#3\}}\right|#2\left|\vphantom{\{#1#2\}}#3\right\rangle }

\newcommand{\curly}[1]{\left\{{#1}\right\}}
\newcommand{\crc}[1]{\left({#1}\right)}

\newcommand{\one}{\mathds{1} }

\newcommand{\dg}{^{\dagger}}

\newcommand{\ee}[1]{\times10^{#1}}
\newcommand{\bb}{\mathcal{B}}
\newcommand{\gb}{\mathcal{G}}

\newcommand{\Schrodinger}{Schr\"{o}dinger}

\newcommand{\pvb}{PvB}
\newcommand{\Bo}{Bi-orthogonal}
\newcommand{\bo}{bi-orthogonal}
\newcommand{\wf}{wavefunction}
\newcommand{\Wf}{Wavefunction}
\newcommand{\wvp}{wave-packet}
\newcommand{\vn}{von Neumann}
\newcommand{\vN}{von Neumann}

\newcommand{\ps}{phase-space}

\begin{document}

\title{Double ionization of Helium from a phase space perspective}

\author{E. Ass\'emat}
\email[]{assemat.elie@gmail.com}
\affiliation{Dept. of Chemical Physics, Weizmann Institute of Science, 76100 Rehovot, Israel }

\author{S. Machnes}
\affiliation{Dept. of Chemical Physics, Weizmann Institute of Science, 76100 Rehovot, Israel }

\author{D. Tannor}
\affiliation{Dept. of Chemical Physics, Weizmann Institute of Science, 76100 Rehovot, Israel }

\date{\today}

\begin{abstract}

The aim of this paper is two-fold. First, we present a phase space perspective on long range double ionization in a one dimensional model of helium. The dynamics is simulated with the periodic von Neumann ({\pvb}) method, an exact quantum method based on a lattice of phase space Gaussians. Second, we benchmark the method by comparing to the Multiconfiguration Time-dependent Hartree method. The {\pvb} approach is found to be faster than the standard MCTDH code for the dynamics calculations and to give better accuracy control.

\end{abstract}



\maketitle

\section{Introduction}

The possibility of controlling electron dynamics in isolated atoms and molecules has generated growing interest in recent years \cite{Review-ARPC-atto,RMP-atto}. Since the timescale for electron dynamics is attoseconds to femtoseconds, to achieve control one would ideally like to have intense fields ($10^{14}-10^{15}$ W/cm$^2$) of several attosecond to subfemtosecond duration. It is now routinely possible to produce NIR fields of $10$ fs duration with intensity on the order of $10^{14}- 10^{15}$ W/cm$^2$ and XUV pulses of femtosecond duration with intensities on the order of $10^{12}$ W/cm$^2$. The desired combination of XUV pulses of sub-fs duration with intensities of $10^{14}-10^{15}$ W/cm$^2$ is still not readily achievable and therefore a slew of recent experiments have employed NIR + XUV pulses, so that the short and weak XUV pulse is boosted by the intensity of the NIR pulse.

From the point of view of simulation, the situation is reversed. The relatively weak XUV pulse, coupling just a few angular momentum states, is much easier to simulate, using e.g. hyperspherical based methods \cite{hypersphe1,hypersphe2} or the R-matrix method \cite{Rmatrix}. In contrast, the intense NIR fields can couple scores or hundreds of angular momentum states, making the simulation extremely expensive or impossible with hyperspherical based methods unless the field intensity in the simulation is reduced well below the usual experimental range of values \cite{hypersphe-NIR}.

Because of the challenge of hyperspherical based simulations, several alternative approaches have been explored. Among them, the Multi-Configuration Time Dependent Hartree method (MCTDH) \cite{MCTDH-paper1,MCTDH-paper2,MCTDH-paper3,MCTDH-paper4} is especially appealing due to its favorable scaling properties. Originally developed to simulate the vibrational dynamics of polyatomic molecules, it has been extended to electronic dynamics as well. It was shown that it can describe the double ionization in a simple model of helium where each electron is described by only one degree of freedom \cite{Scrinzi}. More recently, it was extended to describe double ionization in a two dimensional \cite{Sukiasyan1,Sukiasyan2} and a three dimensional model of helium \cite{helium3d-mctdh} with NIR excitation and two active electrons. In contrast with the hyperspherical approach, MCTDH is able to simulate both strong NIR and XUV pulses.

Several other new methods are currently being developed to describe multielectron dynamics, e.g. the time-dependent generalized-active-space configuration-interaction (TD-GASCI) method \cite{GASCI} and the B-spline algebraic diagrammatic construction \cite{Bsplines}. A recent review of this field with a focus on MCTDH can be found in \cite{review-kiel}.

The above-mentioned calculations have given much insight into the participation of multiple electrons in the ionization process in the presence of strong fields. Despite the significant advances in our understanding, there are several motivations for new computational approaches. 1. Certain aspects of the ionization process may require a much larger spatial range than the one used in the above studies. For example, the study of the time delay in \cite{large-grid-for-time-delay} required a grid more than an order of magnitude larger than the one used in \cite{Sukiasyan1,Sukiasyan2}, and as a result was restricted to the single-active-electron approximation. Note that the t-SURFF method \cite{Scrinzi-tsurff} can describe long range dynamics by construction, but it neglects electron-electron interaction at large distance. The phase space approach presented here, though applied to two electrons in 1-d, is designed to be more efficient than current methods when extremely large grids are required, without neglecting any interaction. 2. The classical picture underlying high harmonic generation by strong NIR pulses is the so-called three-step model consisting of strong field tunneling ionization, quasi-free electron propagation and recollision \cite{Corkum1}. The quantum analog of this three-step model was formulated shortly after the classical model and captures most of the key features \cite{Corkum2}. Normally, full quantum simulation methods do not exploit this underlying classical structure. In contrast, the phase-space propagation method presented here is a fully quantum method that still allows one to see the underlying classical structure of the dynamics and therefore to understand the predictive limits of the classical model. 3. Usually calculations are performed in the coordinate representation despite the fact that some of the key experimental observables are associated with the momentum of the ionizing electrons. An approach that captures the key dynamics simultaneously in coordinate and momentum, i.e. in phase space, could have significant advantages in the analysis of the correlated wavefunction. The advantages of points 2. and 3. could in principle be obtained with other phase space representations, e.g. the Wigner function, but those representations are generally not as compact as the one presented here.

The present approach, called {\pvb}, is based on a periodic lattice of phase space Gaussians, called the von Neumann basis \cite{Asaf-PRL}. The  fact that the method employs a phase space representation enables one to calculate the time evolving wavefunction only in those regions of phase space where there is significant wavefunction amplitude. A by-product of the method is that it gives intuition into the electron dynamics in phase space and the underlying classical correspondence, The {\pvb} method has been applied successfully to the ionization process of one dimensional hydrogen under a combination of strong NIR and XUV pulses \cite{Norio}. Here we extend the method to study the double ionization of two electrons in 1-d in the presence of NIR and XUV pulses. In addition to the multidimentional formulation, a key feature of the method presented here is the algorithm to dynamically adapt the phase space during the quantum dynamics.

The aim of this paper is twofold: first, to present the advantages of a phase space point of view for analyzing double ionization and second, to benchmark the {\pvb} method by comparing it with the MCTDH method. While the MCTDH algorithm is faster to calculate the initial state, for the dynamics calculations presented here, the {\pvb} approach is found to be significantly faster than the standard Heidelberg MCTDH code, as well as to give better accuracy control. However, it is important to mention that this version of MCTDH has not been optimized for ionizing systems and that other implementations of MCTDH may be significantly more efficient. In the future it might be interesting to explore combining the {\pvb} approach with MCTDH to obtain additional computational efficiency.

The paper is organized as follows. In Section 2 we review the periodic von Neumann ({\pvb}) and the MCTDH approaches to simulate quantum dynamics. In Section 3.1 we present the model system which will be used to compare the two approaches. Next we detail the results: a comparison of the performance of the two approaches for eigen-decomposition (Section 3.2) and ionization dynamics, (Section 3.3). Section 3.3 includes a series of phase space pictures of double ionization to highlight the insight given by this approach. Section 4 concludes with a review of the advantages and current limitations of the {\pvb} method and an outlook of how it may be further developed.

\section{Methodology} \label{methodology}

\subsection{The {\bo} {\vn} Basis}

Before going into details we provide a brief summary of the method.  The von Neumann (vN) lattice is a lattice of Gaussians in phase space with one Gaussian per phase space cell of area $h^d$ ($h$ is Planck's constant and $d$ is the dimension) \cite{original-Neumann}.  This basis is known to be complete but not overcomplete on the infinite plane \cite{Perelomov-completeness} \footnote{Technically, the basis on the infinite plane is overcomplete by one.}. The appeal of this basis is that the Gaussians can be placed only where needed in phase space, and hence classical intuition can be used to guide and to interpret the quantum calculation. However, in any calculation on a truncated phase space the vN lattice is known to have severe problems with convergence \cite{davis-heller,daubechies}.  We have shown that by modifying the Gaussians ${g_i}$ to be band-limited and periodic, ${\tilde{g}_i}$, one can obtain Fourier accuracy \cite{Asaf-PRL,{\pvb}-review}.  In order to prune basis functions from regions of phase space that are not necessary one actually needs to use the biorthogonal partners ${b_i}$ instead of the ${g_i}$ as the basis functions as discussed below (for a fuller explanation see \cite{Asaf-PRL,Norio,The-Math-Paper}).  We call this ``biorthogonal exchange", giving rise to the name of the method "periodic von Neumann basis with biorthogonal exchange" or {\pvb}. An alternative method to converge the truncated von Neumann lattice has been developed by Poirier \cite{poirier}.

\subsubsection{The Underlying Hilbert Space}
We begin by defining the Hilbert space which serves as the foundation for all further discussion --- a truncated discrete Fourier basis. As we will see, this finite-dimensional Hilbert space corresponds to a rectangular area in {\ps}.

Choosing a finite region of length $L$ in $x$, we may assume, without loss of generality, cyclic boundary conditions $f\left(x\right)=f\left(x+L\right)$. This, in turn, implies that the functions in
this interval are spanned by the orthogonal functions of the form $\exp\left(2\pi i\frac{x}{L}n\right)=\exp\left(ik_{n}x\right),\,\,\,\forall n\in\mathbb{Z}$,
with $k_{n}=\frac{2\pi}{L}n$. Next, we limit ourselves to $\left|k_{n}\right|\le K_{\textrm{lim}}$,
i.e. $ n \in [-\frac{KL}{2\pi} + 1;  \frac{KL}{2\pi}]$ with $K = \frac{2\pi}{L} \left\lfloor \frac{K_{\textrm{lim}}L}{2\pi}\right\rfloor$, ending up
with a rectangular area of {\ps} spanned by a discrete number,
$N=2 \frac{KL}{2\pi}$, of complex exponential
functions --- a set we shall denote as the \emph{spectral basis}. This Fourier grid of $N$ points spans an area of $N h$ in {\ps} ($L\times2P=L\frac{2Kh}{2\pi}= N h$).

Nyquist's theorem ensures that by sampling the interval at
$N$ uniformly-space points, i.e. at resolution $\delta_{x}=\nicefrac{L}{N}$,
we can fully reconstruct any function residing inside the {\ps}, i.e. the functions spanned by the spectral basis.
Given any two of $L$, $K$ and $\delta_{x}$ (or $N$), we may define the set of sampling points, henceforth denoted the \emph{Fourier grid} \emph{(FG)}.

The set of functions within the Hilbert space that take on the value $0$ at all
grid points except for a single grid point where the value is $1$, is
known as the \emph{pseudo-spectral basis}. These functions can be shown to be orthonormal and span the same Hilbert space as the spectral basis. For
the Fourier grid, the pseudo-spectral functions are the $N$ periodic sinc
functions $\textrm{sinc}_{n}\left(x\right)=\frac{\sin\left(K\left(x-x_{n}\right)\right)}{\sqrt{\pi}K\left(x-x_{n}\right)}$, centered at the $N$ grid points
\cite{David's-Textbook}.

\subsubsection{{\pvb} - {\Bo} {\vn} Basis} \label{def-basis}
\begin{figure}[htbp]
    \begin{center}
        \includegraphics[scale=0.5]{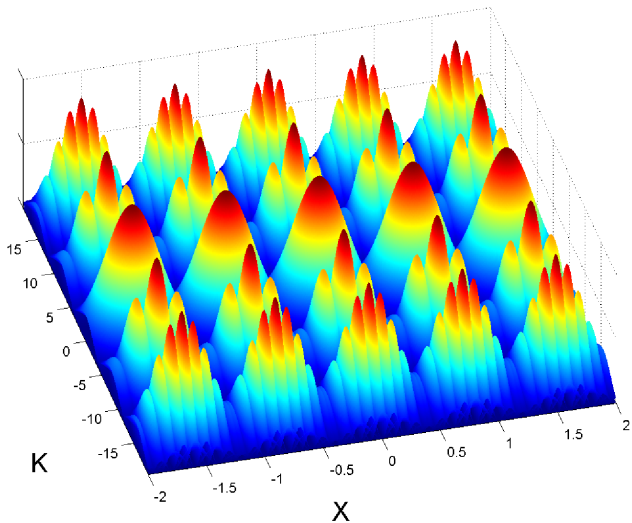}
        \end{center}
    \caption{A $5\times5$ {\vn} grid of Gaussians (real component), spanning the same subspace as a $25$-point Fourier grid, serving as the set of bras for the {\bo} {\vn} basis. The figure depicts a mixed representation showing the oscillations in the basis functions in $x$-space centered at different $k$s.}
    \label{fig:grid_of_Gaussians}
\end{figure}

The Fourier basis described above is capable of giving high accuracy but does not generate a sparse representation of the state. However, given a Fourier basis of $N$ functions, corresponding to a rectangular region of {\ps}, we may span the same space using any set of $N$ linearly-independent functions which are themselves within that subspace. Specifically, to create this linear combination we use a set of phase space Gaussians, centered on a grid of $\sqrt{\alpha N}\times\sqrt{N/\alpha}$ points within the phase space spanned by the Fourier grid. See fig. \ref{fig:grid_of_Gaussians}.

\begin{equation}
g_{nl}(x) = \left(\frac{1}{2 \pi \sigma_x^2}\right)^{1/4} \exp{\left(-\left(\frac{x-x_n}{2 \sigma_x}\right)^2 + \frac{i}{\hbar}p_l(x-x_n)\right)}.
\label{eq:gaussians-def}
\end{equation}

We shall term this grid the \emph{{\vn} lattice} and the basis formed by this set of $N$ Gaussians states as the \emph{{\vn} basis}, with each Gaussian loosely speaking spanning an area of $h$.

Let us denote the set of Gaussian states of the {\vn} basis as $ \gb = \curly{g_{k}}_{k=1}^N$, and their matrix representation in the Fourier grid as the  $N{\times}N$ matrix $G$, with $G_{j,k}=g_{k}\crc{x_j}$, where $k$ is an enumeration of the $N$ Gaussians and $x_j$ are the Fourier Grid points. As the Gaussians are non-orthogonal, we define the \emph{{\bo} basis} to $\gb$, which we shall term $\bb$: $\bb = \curly{b_{k}}_{k=1}^N$ where $\braket{g_{k}}{b_{j}}=\delta_{kj}$. We denote the matrix representation of $\bb$ as the $N{\times}N$ matrix $B$. Note that while $\gb$ is composed of well-localized Gaussians, the functions in $\bb$ are extremely delocalized.

In matrix notation, the biorthogonality takes the form
\begin{equation}
BG\dg=G{\dg}B=I,
\label{eq:mat_biorth}
\end{equation}
i.e. $B={G^{\dagger}}^{-1}$.
We may introduce a vector notation $\psi_B= B^{-1} \psi = G^{\dagger}\psi$, where the elements $\psi_{B_{k}}=\braket{g_{k}}{\psi}$ are the coefficients of $\psi$ in the $\bb$ basis.  We then find that $\psi=B\psi_B=B\crc{G{\dg}\psi}=\crc{BG\dg}\psi$, consistent with eq. \ref{eq:mat_biorth}.

\subsubsection{The {\Schrodinger} Equation}
Let us derive the form of the {\Schrodinger} equation in the {\bo} {\vn} basis. Starting with the standard expressions, $H\psi=E\psi$ and $\partial_{t}\psi=-\frac{i}{\hbar}H\psi$, substituting $\psi=B\psi_{B}$ and shifting the time-independent $B$ to the other side we end up with
\begin{equation}
B^{-1}HB\psi_{B}=\lambda_E\psi_{B}\,.   \label{eq:TISEU}
\end{equation}
Similarly, the time-dependent Schr\"odinger equation becomes
\begin{equation}
\partial_{t}\psi_{B}=-\frac{i}{\hbar}B^{-1}HB\psi_{B}\label{eq:TDSEU}
\end{equation}
Note the appearance of a similarity transform of the Hamiltonian, as opposed to the standard unitary rotation. Also, recall $B^{-1}=G^{\dagger}$.

\subsubsection{The Reduced {\pvb} Basis}

As discussed, in the {\bo} {\vn} basis the representation of localized wavefunctions is sparse. The coefficients that are close to zero can be neglected with minimal and well-controlled loss of accuracy and therefore this representation can be used to propagate dynamics in an efficient manner. Specifically, as $\ket{\psi}=\sum_{k}\psi_{B_{k}}\ket{b_{k}}$ with $\psi_{B_k}=\braket{g_k}{\psi}$,
we expect the $\psi_B$ vector to have a negligible value for areas in {\ps} where $\psi$ is not present.
One may therefore reduce the $\bb$ basis to the subset of $b$-vectors whose coefficients are above some \emph{{\wf} amplitude cutoff} ($W$). Formally, we are projecting the state to a subset of the $\bb$ basis vectors. We shall term this \emph{the reduced basis}. In matrix representation, this corresponds to the selection of some columns of $B$, giving a $N{\times}R$ matrix, with $R{\le}N$, that we denote by $\tilde{B}$. The state is initially represented as a column vector containing $N$ coefficients. After this reduction the state is represented by a column vector of $R$ coefficients such that all of them have an absolute value above $W$. This vector is denoted $\psi_{\tilde{B}}$ and termed \emph{the reduced state}. Note that $R$ can be several order of magnitude smaller than $N$.

As $\tilde{B}$ is no longer square, $\tilde{B}^{-1}$ is not well defined, and it must be replaced in eqs. \ref{eq:TISEU},\ref{eq:TDSEU} by the pseudo-inverse, defined by $\left(\left(\tilde{B}^{\dagger}\tilde{B}\right)^{-1}\tilde{B}^{\dagger}\right)\tilde{B}=I$. This gives
\begin{equation}
  \left(\tilde{B}^{\dagger}\tilde{B}\right)^{-1}\left(\tilde{B}^{\dagger}H\tilde{B}\right)\psi_{\tilde{B}} = \lambda_E\psi_{\tilde{B}}\label{eq:Schrod indep B reduced}
\end{equation}
\begin{equation}
\partial_{t}\psi_{\tilde{B}}=-\frac{i}{\hbar}\left(\tilde{B}^{\dagger}\tilde{B}\right)^{-1}\left(\tilde{B}^{\dagger}H\tilde{B}\right)\psi_{\tilde{B}}. \label{eq:Schrod B reduced}
\end{equation}

The truncation of the $\bb$ basis defines a projection that possesses interesting properties detailed in \cite{The-Math-Paper}. In particular, it was shown in \cite{Porat} that this is the optimal way to project a state in a subspace to obtain a reduced state as close as possible to the original state. The process preserves the form of the $\{b_k\}$ functions, i.e. $\tilde{b}_k=b_k,$ but replaces the Gaussians $\{g_k\}$ by a new set of deformed Gaussians, $\{\tilde{g}_k\}$, that is orthogonal to the reduced set of $\{\tilde{b}_k\}$.

To solve eq. \ref{eq:Schrod B reduced} numerically one can use any standard quantum propagation algorithm able to deal with non-Hermitian Hamiltonians. For this study we use the Taylor propagation presented in \cite{GOAT}, which we found to be the most efficient for our method.

Note that in this formulation, memory and time requirements scale with the reduced basis size, not the full Hilbert space size.

\subsubsection{Finding Eigenmodes in {\pvb}}

We now turn to computing the eigenmodes of a given potential in the reduced {\pvb} basis. We are faced with
a seemingly intractable problem. On the one hand, we are unable to represent the full
Hilbert space in memory due to its unmanageable size, and on the other
hand, we have no foreknowledge of which $b$-vectors make up the appropriate reduced basis. We therefore
utilize an iterative algorithm, detailed in \cite{The-Math-Paper}. The iteration starts at the locations
most likely to be included in the reduced basis --- the potential's local
minima --- and expands the reduced space as needed. We shall refer to regions on the {\vN} lattice at which Gaussians are centered as cells,
allowing us to speak of ``neighboring cells in {\ps}'', ``cells at the boundary of the reduced basis set'', etc.

Following is a pseudocode for the iterative algorithm:

\vspace{5mm}

\label{gs-algo}
\noindent\texttt{10 Define the initial reduced basis consisting of the cells at the position X of the potential's local minima, with $p=0$.}\\
\noindent\texttt{20 Compute eigenmodes for the current reduced basis.}\\
\noindent\texttt{30 If everywhere on the boundary of the reduced basis, all eigenmode amplitudes are below the specified accuracy threshold, stop. If not, continue.}\\
\noindent\texttt{40 Remove all {\ps} cells where the amplitude is below the {\wf} accuracy cutoff.}\\
\noindent\texttt{50 Expand the reduced basis to all neighboring cells (i.e. all cells at or below some distance $r$ from the current reduced basis)}\\
\noindent\texttt{60 Compute the additional elements of the now expanded, reduced Hamiltonian.}\\
\noindent\texttt{70 Go to 20}\\

\vspace{5mm}

\noindent A more detailed description of this, and the following dynamics algorithm, is given in \cite{The-Math-Paper}.

\subsubsection{Dynamics in the Reduced {\pvb} Basis}

The main difference between the dynamics algorithm used in this work and the one used in \cite{Norio}, beyond adaptation to multidimensional systems, is a new dynamic method for choosing the reduced basis set. In \cite{Norio}, the total time was divided in 8 segments. For each of these segments some concatenation of rectangles of phase space was chosen as the reduced basis set based on classical trajectories. Moreover, the dynamics was computed with a fixed time step. Here, the methodology is refined based on the following insight: given that the time evolution of the {\wf} is continuous in the {\ps}, we tailor the reduced basis as the wavepacket evolves in time. We do this by monitoring the {\wf} amplitude at the boundary of the reduced space: If it rises above the specified accuracy threshold, for example at the ``bow'' of a travelling {\wvp}, the reduced basis is expanded in the region of the boundary. Conversely, as the amplitude falls below the threshold at the {\wvp}'s ``stern'', vectors are removed from the reduced basis. Finally, we dynamically adapt the time step to the speed of the wavepacket at the boundary. If the amplitudes at the boundaries grow faster than some speed threshold, we divide the time by two. Conversely, we increase the time step by 20\% if no expansion was needed during the last few time steps. The maximum time step is limited by the rate of change of the field and the propagation scheme, with the actual time step determined by the desired target accuracy.

Note that the new algorithm can describe tunnelling as long as the accuracy is high enough to include the exponentially decreasing wavepacket inside the classically forbidden area.

\subsubsection{Implementation for multidimensional systems}

Consider a Hamiltonian of a system with $N_d$ spatial dimensions that is discretized on a multidimensional Fourier grid:

\begin{equation}
H = \sum_i^{N_d} \one \otimes ... \otimes \frac{p_i^2}{2 m_i } \otimes ... \otimes \one + V(x_1,...,x_{N_d}) \, ,
\label{eq:multidim-ham}
\end{equation}
where $p_i$ are $N$ by $N$ matrices and $V$ is a $N^{N_d}$ by $N^{N_d}$ matrix, with $N$ the number of points on the spatial grid of one dimension. For simplicity assume that the grid has the same number of points in each dimension.

The kinetic part is separable; therefore each 1-d  $p_i$ can be written analytically in the Fourier representation, \cite{David's-Textbook}, then transformed to the unreduced $\mathcal{B}$ basis: $p_i^{B} = B^{-1}p_iB$. Here, as in Section \ref{def-basis}, $B$ denotes the matrix obtained via eq. \ref{eq:mat_biorth} from the matrix $G$ that contains the 1-d discretized Gaussians on the Fourier grid.

Generally, the non-tensor product parts of the Hamiltonian are the
particle-particle interaction terms. We therefore decompose these particle-particle
interactions as
\begin{multline}
V\left(x_{1},x_{2}\ldots x_{N}\right)\\
=\sum_{j_{1}=1}^{m_{1}}\ldots\sum_{j_{N}=1}^{m_{n}}c_{j_{1}\ldots j_{N}}\left(t\right)V_{j_{1}}^{\left(1\right)}\left(x_{1}\right)\otimes\ldots\otimes V_{j_{N}}^{\left(N\right)}\left(x_{N}\right)
\end{multline}
where the single-particle potential vectors, $V^{\left(k\right)}$, are normalized,
and $c$ representing the magnitude of each term. The
general problem of finding an optimal decomposition (with minimal
possible error for any number of terms) is an open problem. For the
purposes of this work we utilize the POTFIT algorithm \cite{POTFIT1,POTFIT2},
which is optimal for two degrees of freedom and applicable generally. Each of the terms $V_{j_{1}}^{\left(i\right)}\left(x_{i}\right)$ is then converted to the $\mathcal{B}$ basis.

To obtain the matrix $\tilde{B}^{\dagger}\tilde{B}$ in eq. \ref{eq:Schrod B reduced}, one first defines the one dimensional set of discretized Gaussians, which gives a matrix $G$, then generates the corresponding one dimensional $B$ matrix by $B = (G^\dagger)^{-1}$ and computes the elements of the two dimensional $B$ matrix defined by the reduced basis $b_{ij} = b_i \otimes b_j$.

At this point, we have the whole Hamiltonian in the $\mathcal{B}$ basis, but stored in memory as a collection of matrices representing 1-d functions. The reduction process of the multidimensional $\mathcal{B}$ basis is then defined on the indices of the complete basis,  $i_{cb}\in [1; N^{N_d}]$. One needs to store in memory a vector which maps the indices of the complete basis to the collection of 1-d indices of the reduced basis, in order to efficiently implement the many changes of reduced basis that occur during the computation of eigenmodes or the dynamics.

\subsubsection{Symmetry and Parallelizability}

Symmetry considerations can cut the number of Hamiltonian elements computed by two orders of magnitude \cite{The-Math-Paper}. Indeed, the $\bb$ basis is a Gabor basis because the $\gb$ is Gabor by construction \cite{gabor1}, which implies that it possesses a translation symmetry in phase space. From the latter, one can deduce that the integral $\int b_{x_m p_a}(x_1,x_2)^*h(x_1,x_2) b_{x_n p_b}(x_1,x_2)$ depends on $(p_a - p_b)$ instead of $(p_a,p_b)$. The exchange symmetry and the hermiticity of the Hamiltonian also reduce the number of elements to compute.

Next, let us remark that {\pvb} is straightforward to parallelize. Indeed, a significant part of the computational effort is spent on converting elements of the reduced Hamiltonian to the {\pvb} basis --- a task which is trivially parallelizable, as each element's computation is independent. In practice, each time the reduced basis is expanded, only the new elements of the reduced Hamiltonian need to be computed. For example computing the ground state on a grid $x\in[-100$ a.u.$;100$ a.u.$]$ and $|K|<15$ for an accuracy threshold $W=10^{-4}$ with six cores, we observe that over $90\%$ of the task is linearly parallelizable. No significant resources have been spent to distribute the tasks between the cores or to merge back the results, which implies that this ratio will not decrease when the number of cores increases. This fact remains valid for a multi-machine architecture. The proportion of workload that is parallelizable is higher for dynamics than for the eigenmode problem, and increases for higher target accuracies, where the reduced Hamiltonian becomes larger. This can be compared, for example, with MCTDH that reaches only around $50\%$ of parallelization for the dynamics on this type of grid because of the contributions from the SPF propagation that is not suited for the parallel MCTDH, as stated in \cite{MCTDH-guide}.

\subsubsection{Filtering of momenta correlations}

Ionization, produced by short NIR or XUV pulses, generates travelling electron wavepackets. The momenta of these packets are actually lower than the maximum momentum component of the ground state wavepacket. Consequently, to analyze momenta correlations one has to remove the bound part of the electronic wavefunction, that would otherwise obfuscate the momenta correlations of ionized wavepackets.

Within the phase space framework, this filtering operation becomes straightforward. Indeed, one just needs to define the phase space volume $\mathcal{V}$ corresponding to the bound states and remove from the reduced state the corresponding $b$-functions such that their orthogonal Gaussians are centered inside this phase space volume. The projector producing the filtering is:
\begin{equation}
P_F = \sum_{k \in \mathcal{V}} \left|b_k\right\rangle \left\langle g_k\right| \, ,
\label{eq:filter-proj}
\end{equation}
where $k \in \mathcal{V}$ means that the center of the $k$-th Gaussian belongs to $\mathcal{V}$.

Here, to obtain the full momenta correlations plot, we define $\mathcal{V}$ by a simple spatial consideration, cutting everything with $|x_i|<30$ a.u. However, a more subtle phase space definition of $\mathcal{V}$ could be use to analyze separate parts of the wavefunction. Phase space filtering includes coordinate space filtering as a special case. For example, one may filter out all single ionization components by removing $b$-functions such that the corresponding Gaussians satisfy the condition ($|x_1|<x_r$) or ($|x_2| < x_r$), with $x_r$ the radius of the atoms. This method does not require any additional computation, contrary to  the one used in \cite{Norio} where many eigenstates need to be computed and then subtracted. 

\subsection{Review of MCTDH}

\subsubsection{Sum-of-Products Ansatz}

Multi-Configuration Time Dependent Hartree (MCTDH, \cite{MCTDH-paper1})
is one of the leading approaches to solving the time-dependent {\Schrodinger}
equation with multiple degrees of freedom. The basic ansatz is as follows. Any multi-particle state
in a finite Hilbert space may be decomposed as a sum-of-products
of \emph{single particle functions }(SPFs), $\phi^{\left(n\right)}$

\begin{multline}
\psi\left(x_{1},x_{2},\ldots,x_{N_d},t\right)\\
=\sum_{j_{1}=1}^{m_{1}}\ldots\sum_{j_{N}=1}^{m_{n}}a_{j_{1}\ldots j_{N_d}}\left(t\right)\phi_{j_{1}}^{\left(1\right)}\left(x_{1},t\right)\ldots\phi_{j_{N_d}}^{\left(N_d\right)}\left(x_{N_d},t\right)
\end{multline}

with $\braket{\phi_{j}^{\left(z\right)}\left(x_{z},t\right)}{\phi_{k}^{\left(z\right)}\left(x_{z},t\right)}=\delta_{jk}$,
$\braket{\phi_{j}^{\left(z\right)}\left(x_{z},t\right)}{\partial_{t}\phi_{k}^{\left(z\right)}\left(x_{z},t\right)}=0$. Each term of this series is called a configuration. Under weak coupling conditions, the particles will be weakly correlated,
and only a few configurations in the above expression will have non-negligible
coefficients, allowing for a very efficient representation of the multi-particle
{\wf}. For larger systems, it is beneficial for the decomposition to
be made hierarchically, to match the hierarchy of system couplings, leading to a multi-layer MCTDH algorithm \cite{MCTDH-multi-layer}. For the purposes of this paper we made use of the original single layer MCTDH implementation package.

To efficiently implement the MCTDH dynamics, one needs to decompose the Hamiltonian as well as the wavefunction
into a sum-of-products form. The  decomposition of the Hamiltonian is needed
by both MCTDH and {\pvb}, as it allows replacing multi-dimensional integrals
with a series of one-dimensional integrals. Such integrations appear in both MCTDH and {\pvb} when converting
the Hamiltonian to the basis of interest.

\subsubsection{Dynamics}

MCTDH dynamics requires equations of motion for both the $a_{j_{1}\ldots j_{N}}$
tensor as well as the SPFs, $\phi^{\left(k\right)}$. Let us start by introducing the following simplified notations:
\[
A_J = a_{j_{1}\ldots j_{N}} ,\quad \Phi_J = \Pi_{k=1}^N \phi_{j_k}^{(k)},
\]
and
\[
 \varphi^{(k)} = (\phi_1^{(k)}, \ldots, \phi_N^{(k)} )^T.
\]
Then, following \cite{MCTDH-paper3} we define the single-hole function
\[
\Psi_l^{(k)} = \sum_{J}A_{j_1 \ldots j_{k_1} j_{k+1} \ldots j_N} \phi_{j_1}^{(1)} \ldots \phi_{j_{k-1}}^{(k-1)} \phi_{j_{k+1}}^{(k+1)} \dots \phi_{j_N}^{(N)}
\]
the mean-fields
\[
\left\langle H\right\rangle_{jl}^{(k)} = \left\langle \Psi_j^{(k)}|H|\Psi_l^{(k)}\right\rangle
\]
and the density matrix
\[
\rho_{ij}^{(k)} = \left\langle \Psi_i^{(k)}| \Psi_j^{(k)}\right\rangle
\]
 Utilizing the Dirac-Frenkel variational principle, one arrives directly at
\[
\partial_{t}A_J=-\frac{i}{\hbar}\sum_L\left\langle \Phi_J | H | \Phi_L\right\rangle A_L
\]

\[
\partial_{t}\varphi^{\left(k\right)}=-\frac{i}{\hbar}\left(1 - \sum_{j=1}^{n_k}\left|\phi_j^{(k)}\right\rangle\left\langle\phi_j^{(k)} \right|\right)\left(\rho^{(k)}\right)^{-1}\left\langle H\right\rangle^{(k)} \varphi^{(k)}
\]
with  $\left\langle H\right\rangle^{(k)}$ the matrix of mean-fields.

In MCTDH, eigenmodes are generally calculated by imaginary-time propagation,
i.e. a relaxation technique. This algorithm has been extended to return multiple eigenmodes
in a single run.

\subsection{Choice of pseudo-spectral basis}

Both {\pvb} and MCTDH require an underlying discrete basis of localized functions to represent the {\wf} at the grid points.
More precisely, MCTDH uses exclusively one-dimensional grids that support the SPFs whereas the {\pvb} lattice covers the multidimensional {\ps} but is never entirely used.
While many possible pseudo-spectral bases are possible for MCTDH and {\pvb}, as shown in \cite{MCTDH-paper3} and \cite{{\pvb}-review} respectively, we have chosen the Fourier pseudospectral functions for {\pvb} and the harmonic oscillator pseudospectral functions for MCTDH for the sake of simplicity.

\subsection{Scaling of {\pvb} vs. MCTDH}\label{scaling-paragraph}

Let us compare the numerical effort required for {\pvb} vs. MCTDH. With MCTDH, the effort has two terms, one from the basis function evolution and one from the calculation of the coefficients: $N_t(mndN^2 + md^2n^{d+1})$\cite{MCTDH-paper3}, where $N_t$ is the number of time steps, $m$ is the number of terms in the POTFIT series, $n$ is the number of single particle functions in each dimension, $d$ is the dimension and $N$ is the number of points in the one dimensional grid. For low dimensional problems on large grids the first term $N_tmndN^2$ dominates. With {\pvb} the numerical effort decomposes into two parts. The first is the cost to precompute the Hamiltonian terms in the {\pvb} basis,  $mN_rN^2 + N_rN^3$, where $N_r$ is the number of functions in the reduced basis. The first term $mN_rN^2$ corresponds to the conversion of the POTFIT series and the second term $N_r N^3$ comes from the conversion of the kinetic Hamiltonian. The second part of the effort is the dynamics itself, scaling as $N_tN_r^2$. Note that since a large number of timesteps $N_t$ is required to adequately represent the control field, the term proportional to $N_t$ dominates for the calculations in this paper.  For instance in the example used in Section IIIC, $N_t = 15000$ and $5000 \leq N_r \leq 10000$ while $N=4000$, $n=12$ and $m\approx N$. Considering just the term proportional to $N_t$, this translates to a two order of magnitude speedup of {\pvb} as compared to MCTDH, including the cost of the conversion to the {\pvb} basis. In the case where the conversion results are already available from a previous run, the speedup increases to four orders of magnitude. This difference in speed between the two methods increases with the size of the grid.

Note that if the POTFIT series were shorter, MCTDH dynamics would scale very similarly to {\pvb} dynamics. Thus, this large advantage of {\pvb} is valid only for potentials that are difficult to decompose, which is the case for the two-electron Coulomb potential as explained below.

\section{Application to 1-D Helium} \label{results}

\subsection{The Model System} \label{results-model}

To test the suitability of {\pvb} and MCTDH for simulations that combine both bound and unbound dynamics, we apply
 both methods to the double ionization of 1D helium.
We shall begin with the calculation of eigenmodes, and then proceed to the ionization dynamics.

\begin{figure}[htbp]
\begin{center}
\includegraphics[scale=0.5]{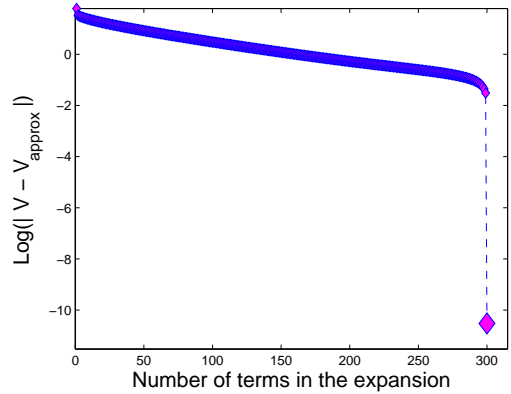}
\end{center}
\caption{POTFIT: Sum-of-products representation for the Coulomb
potential ${V_{approx}}$ on a grid $x\in[-100,100]$ with $300$ points compared to the exact potential $V$. Note that the removal of even a single term from the expansion representing the Coulomb
decomposition leads to a significant drop in accuracy. \label{fig:POTFIT series length}
}
\end{figure}

Our benchmark system, the 1D helium model, consists of two
electrons, each with a single degree of freedom, interacting with
each other and a central (nuclear) potential.
 We use a regularized form
of the Coulomb potential, $\frac{1}{\sqrt{r^{2}+a_{0}^{2}}}$, where
the regularizer, $a_{0}$, is set to $0.739707902$, such
that the ground-state energy of the model matches the experimentally measured binding
energy of helium, $2.903385\,\textrm{ amu}$ \cite{NIST-spectroscopic-data,NIST-fund-const}. For the purpose of high accuracy benchmarking, some of the following results take as reference a ground state energy $E=-2.90338599$ where only the first six decimals are experimentally relevant.

The Hamiltonian used is therefore:
\begin{multline}
H=\frac{1}{4\pi\epsilon_{0}}q_{e}Q\left(\frac{1}{\sqrt{x_{1}^{2}+a_{0}^{2}}}+\frac{1}{\sqrt{x_{2}^{2}+a_{0}^{2}}}\right)\\
+\frac{1}{4\pi\epsilon_{0}}\frac{q_{e}^{2}}{\sqrt{\left(x_{1}-x_{2}\right)^{2}+a_{0}^{2}}}+\frac{1}{2m_{e}}\left(p_{1}^{2}+p_{2}^{2}\right)\label{eq:He1D Hamiltonian}
\end{multline}
with $Q=-2q_{e}$, $q_{e}$ being the electron charge. Note that the interaction term, $\frac{1}{\sqrt{\left(x_{1}-x_{2}\right)^{2}+a_{0}^{2}}}$, is notoriously difficult to represent as a sum-of-products,
requiring a very large number of elements in the series to achieve
an accurate representation (see fig. \ref{fig:POTFIT series length}).

\subsection{Ground State} \label{results-gs}

\subsubsection{Computational Speed and Memory}
We begin by comparing the relative efficiency and behavior of MCTDH and {\pvb} when solving the time independent {\Schrodinger} equation. In the following discussion the range $x\in [-100$ a.u. ; $100$ a.u.$]$ is used for calculating the ground-state.

\begin{figure}
\centering
    \begin{minipage}{.45\textwidth}
        \centering
        \includegraphics[scale=0.4]{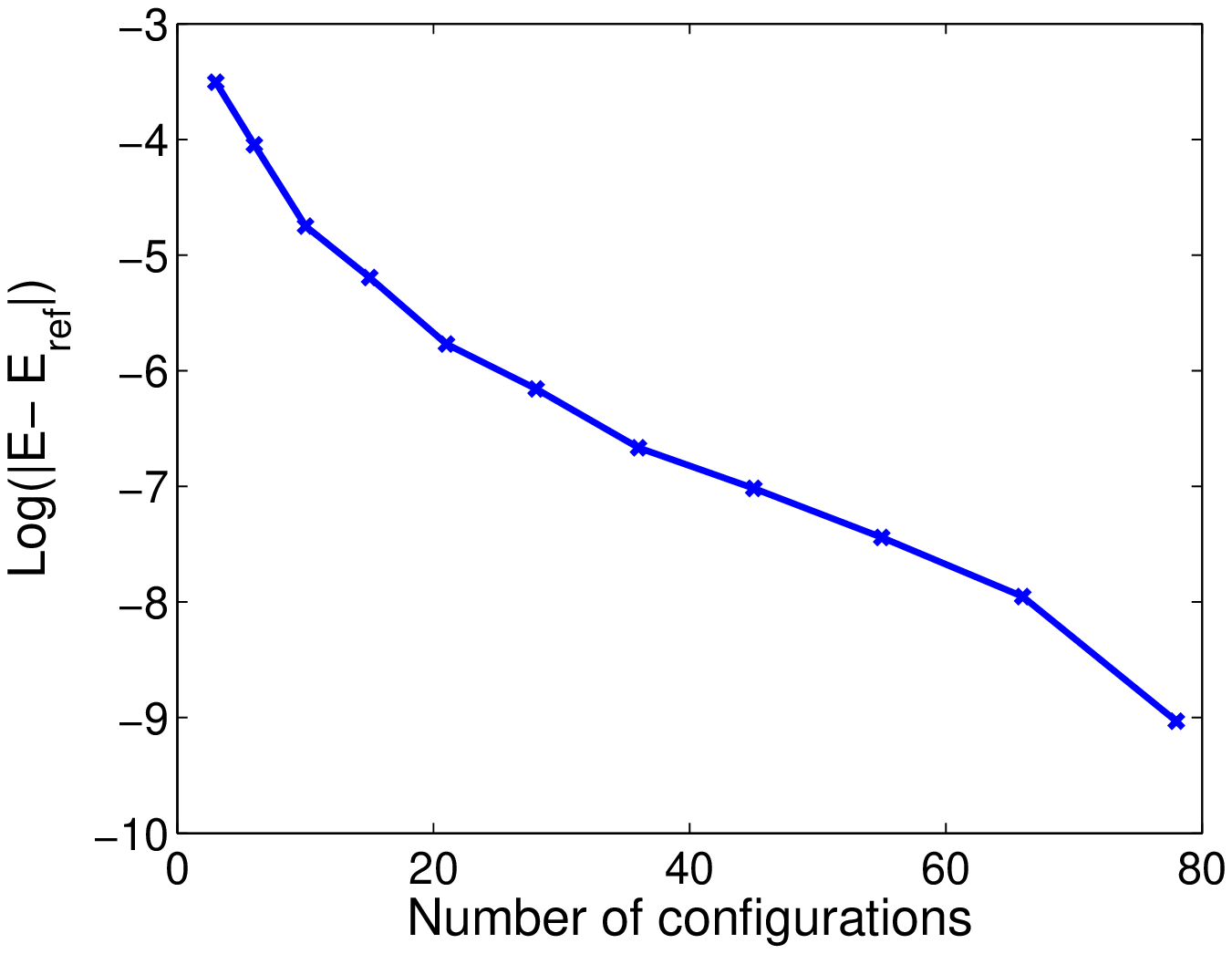}
        \caption{Convergence of the MCTDH ground state energy as a function of number of configurations.}
        \label{fig:MCTDH-convergence-with-nb-config}
    \end{minipage}
    \,
    \begin{minipage}{.45\textwidth}
        \centering
        \includegraphics[scale=0.4]{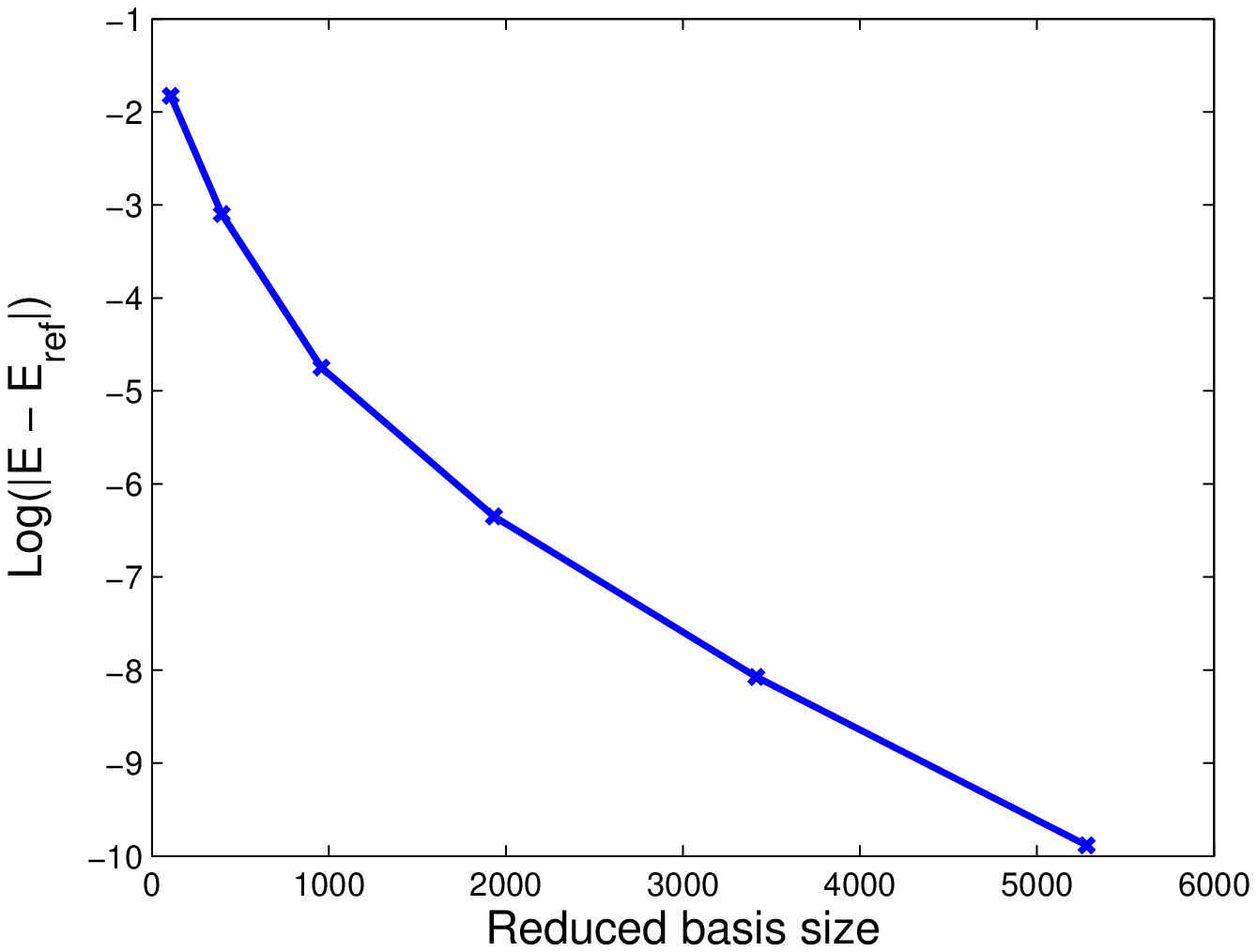}
        \caption{Convergence of the {\pvb} ground state energy as a function of reduced basis size.}
        \label{fig:convergence-with-reduced-size}
    \end{minipage}
\end{figure}

We used the so-called improved relaxation method implemented in MCTDH to solve the eigenvalue problem.
The MCTDH accuracy improves rapidly with the number of configurations, as shown in fig. \ref{fig:MCTDH-convergence-with-nb-config} where $80$ configurations are enough to reach $10^{-9}$ accuracy for the ground energy. Note that we consider here the total number of configurations and not the number per dimension. {\pvb} converges more slowly as a function of the basis size: for the same accuracy $3500$ cells are required, as can be seen in fig. \ref{fig:convergence-with-reduced-size}. Thus MCTDH is generally faster than the {\ps} algorithm by an order of magnitude. However it requires a tuning of the initial guess, whereas the {\ps} algorithm only needs to choose the grid size and the desired accuracy. Moreover, it is interesting to compare how many complex numbers are needed to store the state in memory. In the case of MCTDH, it corresponds to the number of configuration times the size of the one dimensional grid, which gives here $80\times 300 = 24000$, whereas for {\pvb} it is given directly by the reduced basis size: $3500$ in this case. Therefore, for low dimensional problems, the representation of the state is more efficient in the {\pvb} method.

\subsubsection{Accuracy Control}
One of the distinct advantages of the {\ps} algorithm is its accuracy control. This is in contrast to MCTDH, which is unable to detect when the number of grid points is insufficient. Several runs with different grid resolutions are necessary to ensure convergence. If the number of points is too small, MCTDH's improved relaxation will converge with machine accuracy towards a wrong eigenvalue. In contrast, the {\ps} algorithm knows intrinsically which accuracy it achieves.

To illustrate this point, consider a grid with $N=100$ points in each dimension, with $x\in[-20$ a.u. ; $20$ a.u.$]$. This corresponds to a maximal $k$ of $k_{\textrm{max}}=N\pi/L = 7.85$. On this grid, MCTDH returns a ground-state energy of $E = -2.903379690969$, where the improved relaxation converged up to the last digit of this value. On the other hand, the {\pvb} algorithm halts with an accuracy estimate of $10^{-4}$, because it detects that the ground state has reached the edge of the {\ps} area. This accuracy is in fact the real accuracy, as can be verified by running the MCTDH improved relaxation with $N=255$ points, which produces $E=-2.90338601$.

\begin{figure}[htbp]
\begin{center}
\includegraphics[scale=0.3]{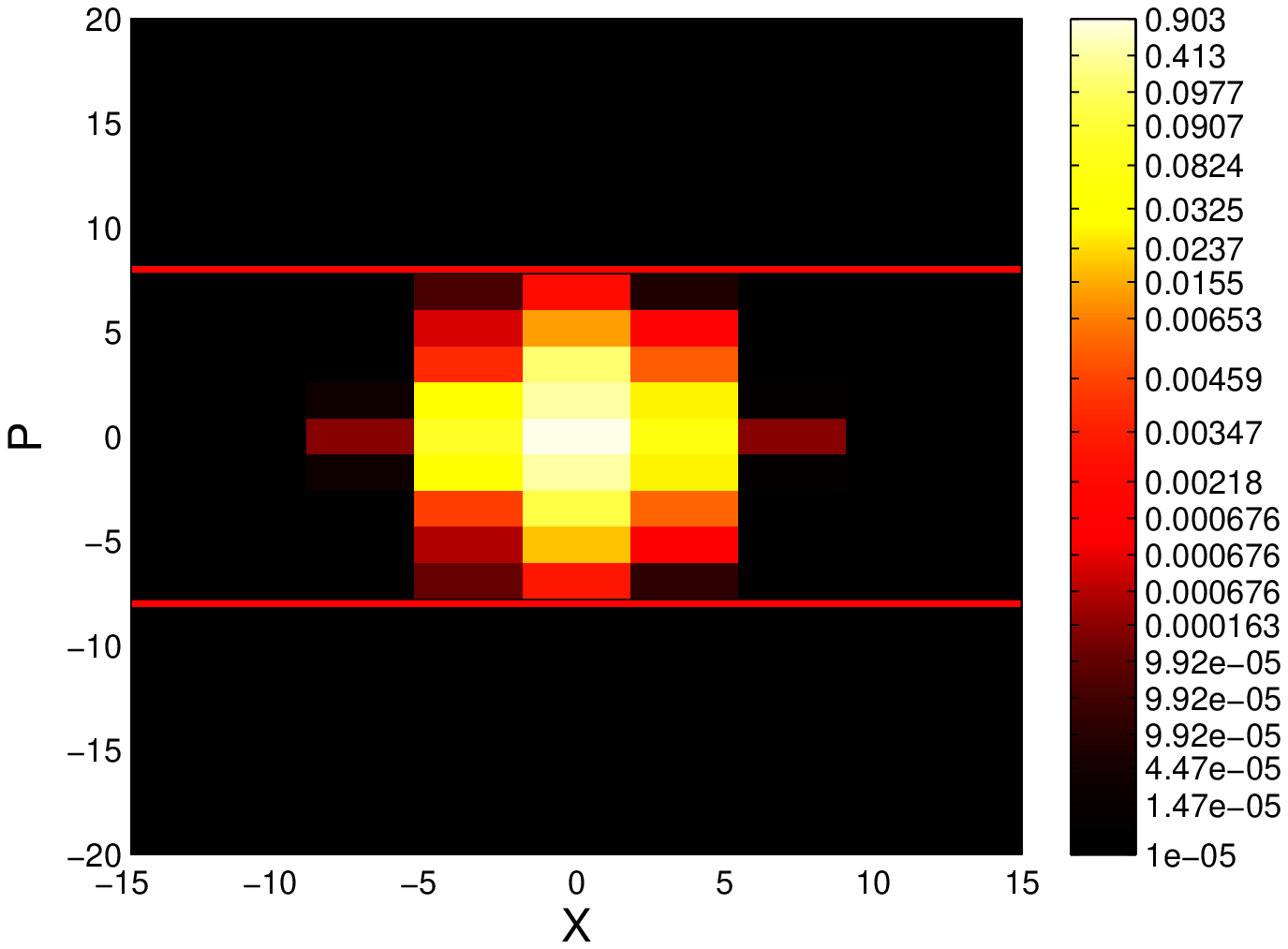}
\includegraphics[scale=0.3]{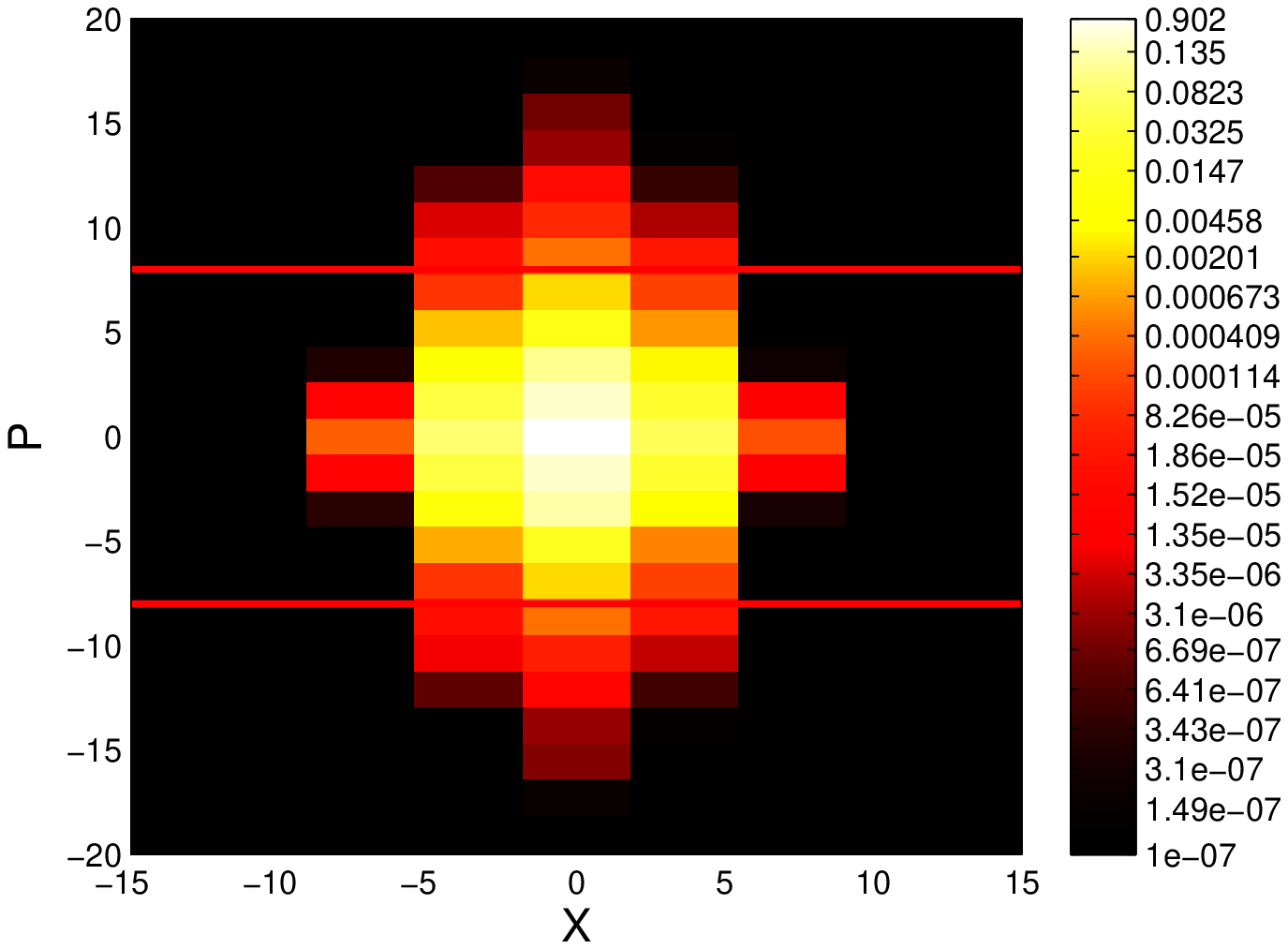}
\end{center}

\caption{Comparison of the one dimensional {\ps} projections of the ground state, obtained with $N=100$ grid points (upper) and $N=255$ (lower). }
\label{fig:groundstatecomparison}
\end{figure}

Insight into the problem can be obtained by inspecting the {\ps} representation of the state. Figure \ref{fig:groundstatecomparison} shows the one dimensional projection (partial trace followed by summation of absolute value amplitudes) of the MCTDH final state represented in the {\pvb} basis. Each unit square represents the position of a Gaussian in {\ps} and the color map represents the amplitude of the overlap between this Gaussian and the state of interest. The upper and lower plots correspond respectively to $N=100$ and $N=255$. The red lines depict the limits of the {\ps} in the case $N=100$. In the case $N=255$ the state wavenumber goes well beyond the $N=100$ limit, which is another way to say that $N=100$ has insufficient resolution to describe the high frequency components of the ground state, leading to inaccurate results. For a large grid, e.g for ionization problems, the long running times make it important to avoid repeating the same computation with different grid sizes to ensure convergence.

The {\pvb} algorithm gives not only an a posteriori evaluation of the accuracy; it allows an a priori estimate. This estimate depends only on the {\Wf} Amplitude Cutoff, $W$, which determines which cells are included in the reduced basis, based on the value of $\left|\braket{g_k}{\psi}\right|$. Consider the reduced state and its complement \cite{The-Math-Paper}
\begin{equation}
\left|\tilde{\psi}\right\rangle = \sum_{k=1}^{N_R}c_k \ket{b_k} \quad \textrm{and} \quad \left|\bar{\tilde{\psi}}\right\rangle = \sum_{k=N_R+1}^{N}c_k \ket{g_k}\,.
\label{eq:eq:energyWAC1}
\end{equation}
The state then decomposes to $\ket{\psi} = \ket{\tilde{\psi}} + \ket{\bar{\tilde{\psi}}}$. Assuming $\ket{\psi}$ is an eigenstate of energy $E$, we can quantify the contribution of the two sets to the energy:
\begin{align}
E & = \Braket{\psi}{H}{\psi} \nonumber\\
& = E^{(0)} + 2E\braket{\bar{\tilde{\psi}}}{\bar{\tilde{\psi}}} - \Braket{\bar{\tilde{\psi}}}{H}{\bar{\tilde{\psi}}}  \nonumber \\
& = E^{(0)} + \sum_{j,k=1}^{N_w} c_j^* c_k \left(2E\braket{g_j}{g_k} - \Braket{g_j}{H}{g_k}\right)
\label{eq:energyWAC2}
\end{align}
where $E^{(0)} = \Braket{\tilde{\psi}}{H}{\tilde{\psi}}$ and $N_w$ is the number of elements in the complement set that are close enough to the border of the reduced set to have non negligible coefficients. We have also made use of the fact that $\braket{\psi}{\bar{\tilde{\psi}}} = \braket{\bar{\tilde{\psi}}}{\bar{\tilde{\psi}}}$, as the reduced set and its complement are orthogonal. Note that there is no first order term in $c_k$, indicating that a perturbation to the eigenstate does not contribute to the energy to first order. This is as expected from the Rayleigh-Ritz variational principle.

By construction, the coefficients in the second term satisfy $c_k\leq W$. Thus, the energy error is bounded by a term of the form $\left|E - E^{(0)}\right| \leq C N_w W^2$, where the constant $C$ bounds the term $\left(2E\braket{g_j}{g_k} - \Braket{g_j}{H}{g_k}\right)$.

From these considerations, we can deduce a heuristic bound on the error for the 1D helium model. We first note that $N_w$ itself depends on $W$ because a larger reduced basis is needed to run a lower $W$ computation. For the 1D helium model we observe in the numerical simulation that $N_w$ scales approximately as $W^{-1/2}$. Thus, the overall bound error bound on  $\left|E - E^{(0)}\right|$ scales approximately as $W^{3/2}$, as shown in fig. \ref{fig:{\pvb} energy as f(WAC)}. A systematic study would be required to validate this bound for others systems, however $W$ itself can still be used as a conservative estimate of the error in other systems.

\begin{figure}[htbp]
\begin{center}
\includegraphics[scale=0.5]{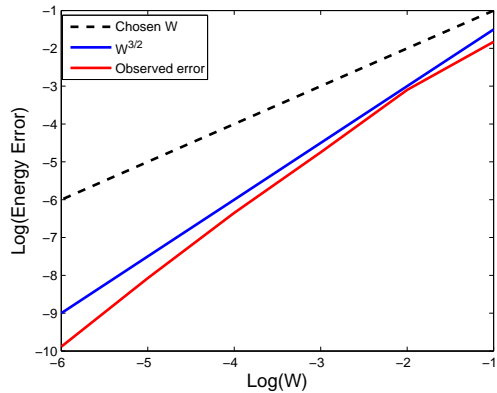}
\end{center}
\caption{Error in the ground state energy as a function of the {\wf} amplitude cutoff $W$.}
\label{fig:{\pvb} energy as f(WAC)}
\end{figure}

\subsection{Dynamics} \label{results-dyn}

\begin{figure}[htbp]
\begin{center}
\includegraphics[scale=0.6]{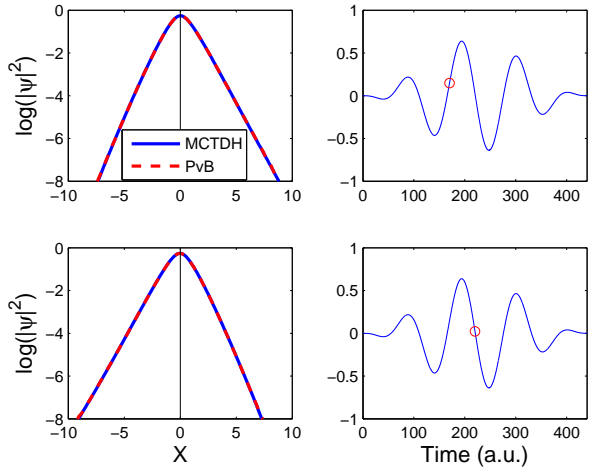}
\end{center}
\caption{ Comparison of the one dimensional projection of the dynamic state computed with {\pvb} and MCTDH under the action of the NIR pulse at two different times (top and bottom rows). Note how the center peak of the {\wf} is tilted left and right, depending on the phase of the driving field, (the red dot in the right panel). Also note the excellent match between the two methods.
}
\label{fig:compare_dyn}
\end{figure}

We now proceed to the dynamics calculations, performing all calculations in the velocity gauge. We first validate the {\pvb} code by comparing to MCTDH on a small grid. Figure \ref{fig:compare_dyn} shows the electronic wavefunction at different times under the influence of just an NIR pulse. In the velocity gauge, the controlled Hamiltonian takes the form:
\begin{equation}
H = H_0 - \frac{q_e}{m_e}(u_{\textrm{\textrm{NIR}}}(t) + u_{\textrm{XUV}}(t))(p_1 + p_2)\, ,
\label{eq:control_ham}
\end{equation}
where $H_0$ is the drift Hamiltonian (eq. \ref{eq:He1D Hamiltonian}) and $u$ is the amplitude of the electric potential. For this first test, the XUV pulse is set to zero. The NIR pulse is taken to have a sine envelop  in order to have exactly zero derivatives at the beginning and at the end: $u_{\textrm{\textrm{NIR}}} = A_{\textrm{\textrm{NIR}}} \sin(2\pi t/T_{\textrm{\textrm{NIR}}} - \pi) \sin(\pi t /(4T_{\textrm{\textrm{NIR}}}))^2$ with $T_{\textrm{\textrm{NIR}}} = 110.32$ a.u. ($= 2.6685$ fs). This corresponds to a wavelength of $800$nm, and a total duration of $10.67$fs. The peak amplitude is $A_{\textrm{\textrm{NIR}}} =0.6627$ (corresponding to an intensity of $5\ee{13} \textrm{W/cm}^2$).

The computation is carried out for an $x$ range $[-100\ldots 100]$ and $N=1000$ points in each dimension. Note that what is shown is a one dimensional projection on the Fourier grid of the two dimensional state. Although the {\pvb} state is pixelated, the projection onto the Fourier grid is smooth, since the {\pvb} state contains by construction exactly the same information.

\begin{figure*}[htbp]
\begin{center}
\includegraphics[scale=0.5]{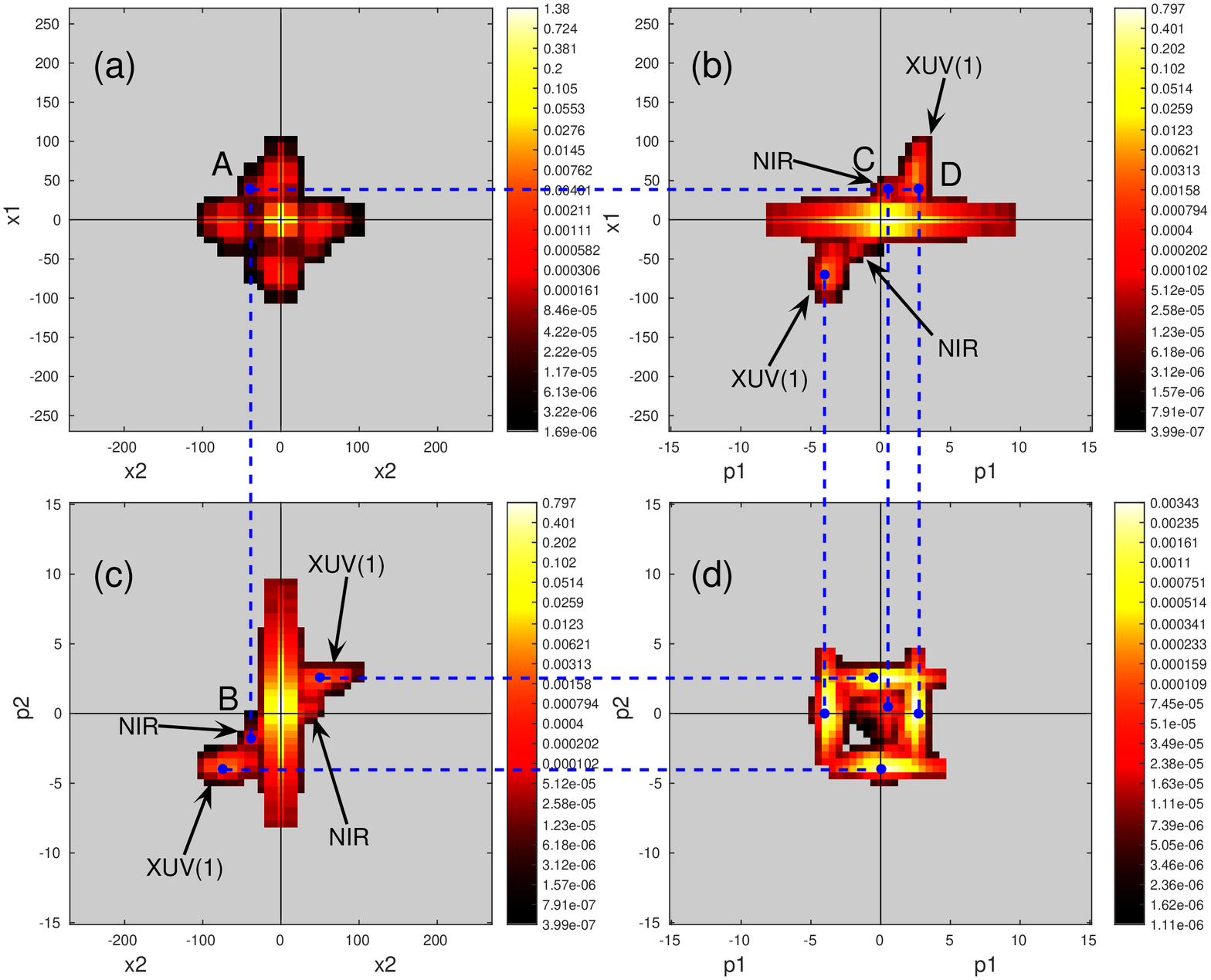}
\end{center}
\hspace{-1.4cm}\includegraphics[scale=0.5]{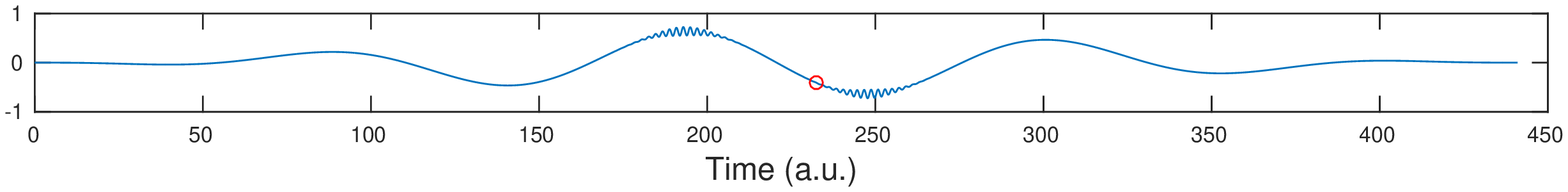}
\caption{Two electron wavefunction before the second XUV pulse at $t=232$ (a.u.). A double ionization marked by (A) is generated by the influences of NIR pulse, point (B) and point (C), and the contribution of the wavepacket ionized by the first XUV, marked by (D)}
\label{fig:Evolution-frames2}
\end{figure*}

\begin{figure*}[htbp]
\begin{center}
\includegraphics[scale=0.8]{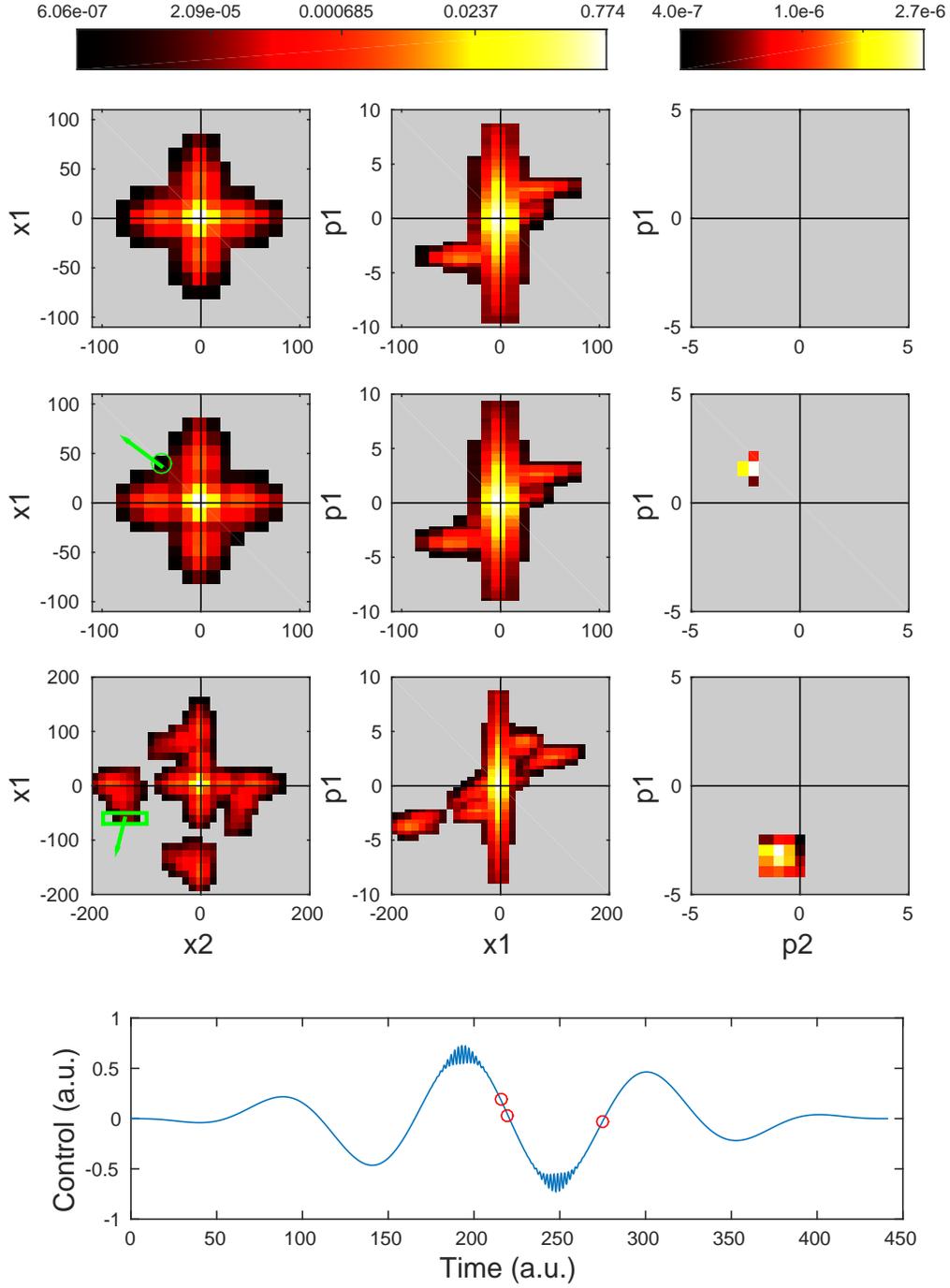}
\end{center}
\caption{Snapshots of the two electron wavefunction between the two XUV pulses, at times $t=215.6$ a.u., $t=219.7$ a.u. and $t=274.7$ a.u., from top to bottom. The first column shows the $x_1-x_2$ correlations. The second column shows the one dimensional phase space projection of the first electron. The third column shows the filtered $p_1-p_2$ correlations, see text for details. The times corresponding to the three snapshots are indicated by red circles on the control pulse in the bottom frame. }
\label{fig:snapshots}
\end{figure*}

\begin{figure*}[htbp]
\begin{center}
\includegraphics[scale=0.5]{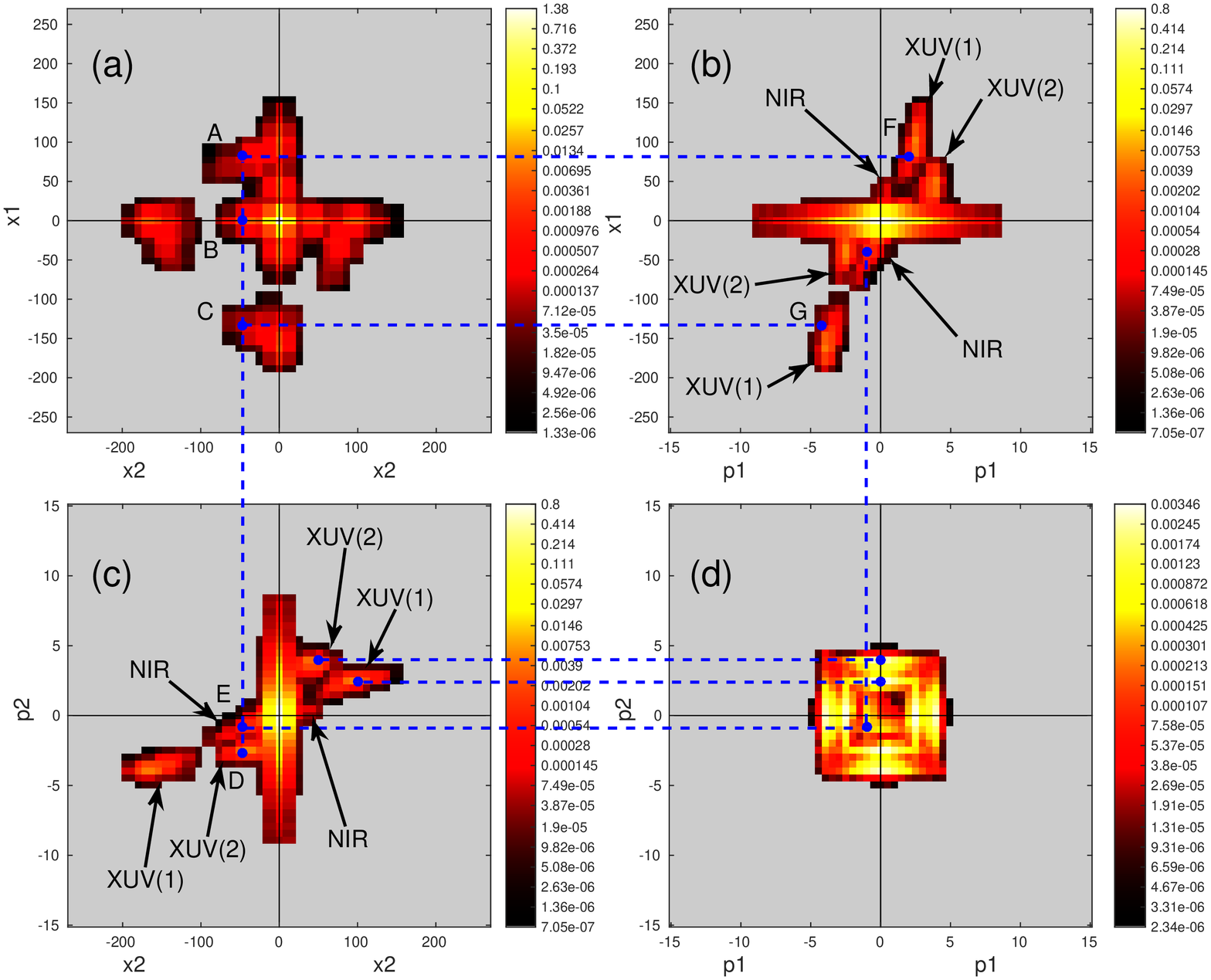}
\end{center}
\hspace{-1.4cm}\includegraphics[scale=0.5]{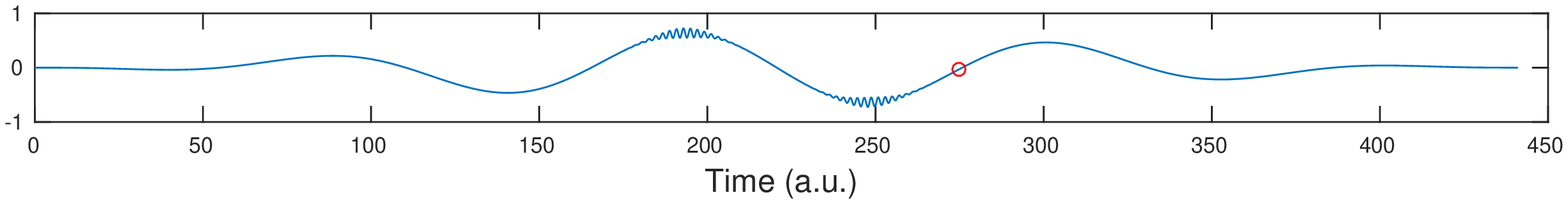}
\caption{Two electron wavefunction after the second XUV pulse at $t=275$ (a.u.). A sequential double ionization marked by (A) and (C) is generated by the influence of the second XUV pulse marked by (D) and the NIR marked by (E), which correlates with the first ionization marked by (F) and (G). The second XUV also generates another single ionization wavepacket, marked by (B).}
\label{fig:Evolution-frames3}
\end{figure*}

\begin{figure*}[htbp]
\begin{center}
\includegraphics[scale=0.45]{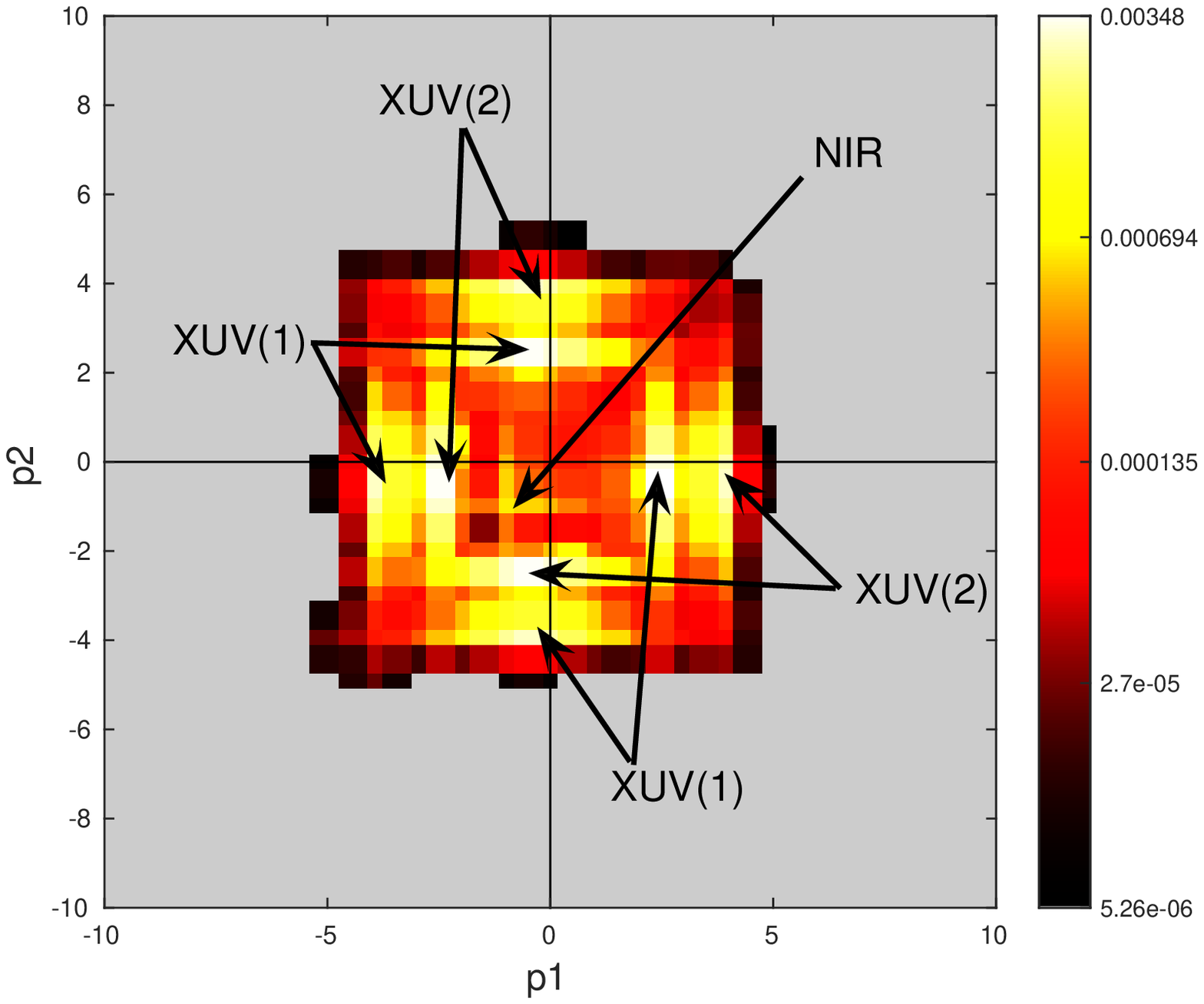}
\end{center}
\caption{Photoelectron momentum distribution at the end of the pulse.}
\label{fig:spectrum}
\end{figure*}

In order to obtain a rich ionization dynamics, we add to the NIR pulse two XUV pulses. While it has long been known that helium can be ionized by a strong NIR pulse, the combination of an NIR pulse  to excite the atom to high bound states, in conjunction with one or more XUV pulse triggering the ionization, allows for much better control of the resulting dynamics. We based these pulses on the one used in \cite{Norio} which has the form: $u_{\textrm{XUV}} = A_{\textrm{XUV}}\sin(2\pi t/T_{\textrm{XUV}} )\exp(-(t - 5T_{\textrm{XUV}}/4 )^2/(2\sigma^2))$ with $T_{\textrm{XUV}} = 2.07$ a.u. which corresponds to a wavelength of $15$ nm. We take $\sigma = 6.207$ a.u. ($150$ attoseconds) and peak amplitude $A_{\textrm{XUV}} = 0.00176$ a.u. (corresponding to an intensity of $=1\ee{12} \textrm{W/cm}^2$). However, to obtain a large amount of double ionization and make the qualitative analysis clearer we increased the amplitude of the two XUV pulses by a factor of $50$. The complete control pulse is shown in fig. \ref{fig:snapshots}. The control strategy is to generate a sequential double ionization with the two successive XUV pulses. For this simulation the {\pvb} calculation is based on an underlying grid with a range $x\in[-400;400]$ with $N=4000$ points in each dimension. On such a grid, the dimension of the unreduced Hilbert space is $16 \times 10^6$ while the maximum dimension of the reduced Hilbert space used during the dynamics is $28207$. This translates to a reduction by 5 orders of magnitude in the number of elements in the Hamiltonian, as compared to the size of the unreduced Hamiltonian.

With these settings, MCTDH is typically two orders of magnitude slower than {\pvb}. The difference in speed is a result of the large number of terms required in the POTFIT series when dealing with the long range multielectron Coulomb potential. For this range, the number of terms in the POTFIT series becomes on the order of $N$ (see fig. \ref{fig:POTFIT series length}), which drastically reduces the efficiency of MCTDH, as explained in subsection \ref{scaling-paragraph}.

The two electron dynamics are shown in figs. \ref{fig:Evolution-frames2}, \ref{fig:snapshots} and \ref{fig:Evolution-frames3}. Figures \ref{fig:Evolution-frames2} and \ref{fig:Evolution-frames3} have the following structure. Frame (a) is the wavefunction projected onto the $(x_1,x_2)$ plane. Single ionization corresponds to the wavepacket moving along the horizontal and vertical axes of the frame (a) while double ionization corresponds to the wavepacket moving along the diagonals of the same frame. The frames (b) and (c) show the phase space projection of the two electrons. These frames are identical due to symmetry, up to a switching of axes to facilitate the reading of the correlations.  Frame (d) presents the wavefunction projected onto the $(p_1,p_2)$ plane, after being filtered to remove the contribution from the bound states. The plots are produced by a 2-d projection of the 4-d phase space amplitude, e.g. the amplitude of one cell $|C| = \left|\left\langle g(x_1^{(k)},p_1^{(m)},x_2^{(l)},p_2^{(n)})|\psi\right\rangle\right|$ in the spatial correlations plane is:
\begin{equation}
\left|C_{\left(x_1^{(k)},x_2^{(l)}\right)}\right| = \sqrt{\sum_{m,n} \left|C_{\left(x_1^{(k)},p_1^{(m)},x_2^{(l)},p_2^{(n)}\right)}\right|} \, ,
\label{eq:2d-projection}
\end{equation}
where $(x_1^{(k)},p_1^{(m)},x_2^{(l)},p_2^{(n)})$ denotes the centers of the Gaussians. This choice may not be the best as it induces a strong loss of information but is is straightforward to implement numerically and efficient enough for the purpose of analyzing the position of the wavepackets in phase space.

Figure \ref{fig:Evolution-frames2} shows the projections at $t=232$ a.u.,  before the second XUV pulse. Inspection of the wavepacket dynamics reveals that the amplitudes marked by B and C are generated by the NIR pulse while the amplitude labelled by D has been generated by the first XUV pulse. This labelling is confirmed by checking the momentum expected from the frequency of the pulses with the de Broglie relation, which gives $p_{XUV}\approx 3.04$ and $p_{\textrm{NIR}} \approx 0.05$. Of course, one should not expect to observe exactly these values, since the dynamical process is complicated, but they allow one to discriminate between travelling wavepackets with a low $|p|$ and higher $|p|$, produced respectively by the NIR and XUV pulses. Note that the projections (dashed blue lines) in frames (b) and (c) allows one to distinguish the different contributions to the double ionization wavepacket. The projection lines show that the double ionization amplitude may include contributions from both pulses. Frame (d) shows the expected momenta of the different wavepackets.
Note that after the bound amplitude is filtered out all momenta are lower than the maximum momentum of the bound states, that are not filtered out in frame (b) and (c).

Figure \ref{fig:snapshots} shows three snapshots to emphasize the presence of concerted and sequential double ionization. The middle column presents the same projection as frame (c) of figures \ref{fig:Evolution-frames2} and \ref{fig:Evolution-frames3}. The left column plots the ($x_1-x_2$) correlations, while the right column plots the ($p_1-p_2$) correlations filtered such that only the double ionization is visible. On the first line, only single ionization is present, thus the right plot on this line is empty. The second line shows the first appearance of double ionization ($t=219.7$), which corresponds to only one cell in the first column, indicated by the green circle at $x_1 \approx x_2$. One can see in the right column that both momenta are nonzero, and approximatively equal. The combination $x_1 \approx x_2$ and $p_1 \approx p_2$ is the signature of concerted double ionization. Thus, there is a small amplitude of concerted ionization at this point in time. The double ionization becomes greater at later time and largely sequential, as shown in the third line ($t=274.7$). The green rectangle in the left frame of the third line is located at $\left\|x_2\right\| \gg \left\|x_1\right\| $. The right frame of the third line shows the filtered momentum correlations corresponding to that rectangle. We observe the wavepacket, originating at $x_1 \approx 0$, $x_2 < 0$, has momentum $p_2 \approx 0$ with a significant negative $p_1$. This is the signature of sequential double ionization. The green arrows on the first column show the direction of the motion obtained from the average of the filtered wavepackets shown on the third column.

Figure \ref{fig:Evolution-frames3} shows the projections at $t=275$ a.u., following the second XUV pulse. The second XUV pulse generates a significant amount of sequential double ionization (fig. \ref{fig:Evolution-frames3}a, points A and C) as well as a second wave of single ionization (fig. \ref{fig:Evolution-frames3}a, point B). The double ionization in this case is sequential, because the corresponding wavepackets arrived at their present location by a two step process, first following the axis before leaving it to get closer to the diagonal and anti-diagonal of fig. \ref{fig:Evolution-frames3}a.  The NIR pulse has the effect of shifting the new single ionization wavepacket toward positive momentum compared with the single ionization from the first XUV. Indeed, it can be seen in fig. \ref{fig:Evolution-frames3}c that the two lobes with higher $|x_2|$ generated by the first XUV pulse have lower momentum than the one generated by the second XUV pulse. Note again that the dashed blue projection lines allow one to determine by eye the contributions to the double ionization amplitude. For example the amplitude marked C corresponds to a superposition of amplitude from G, generated by the first XUV pulse, from E, generated by the NIR pulse and from D, generated by the second XUV pulse.

Figure \ref{fig:spectrum} shows the photoelectron momentum distribution at the end of the pulse. We observe the expected contributions from the different parts of the pulse. The correlations are bounded by the momenta of highest absolute value, corresponding to the XUV pulse. For a given electron, each of these high momentum wavepackets correlates with all momenta of the other electron, which creates the square shape.  One can also notice the ponderomotive shift between the two XUV wavepackets, for example for $p_2 = 0$ and $p_1>0$. This shift is approximatively 2 atomic units, which corresponds to the NIR total amplitude. Note that the drop in intensity of the boundary layer comes from the phase space truncation, these cells having some contributions from the core cells that were filtered out.

Finally, note that the accuracy of this simulation is $10^{-8}$ for $|\psi|^2$, ($W=10^{-4}$), which is similar to many other studies, see for example \cite{Sukiasyan2}.

\section{Conclusion} \label{conclusion}

We have applied the {\pvb} method to the calculation of both eigenstates and ionization dynamics. The ground state calculation (and eigenstate calculations in general) requires a relatively small grid, and in such cases {\pvb} is slower than MCTDH. However, we have shown that {\pvb} offers better control of the accuracy. On the dynamics side, we showed that {\pvb} and MCTDH reproduce the same short-time dynamics, but for the large and dense grids required for the long-time dynamics {\pvb} is faster than the Heidelberg MCTDH code by orders of magnitude. Indeed, it can deal more easily with the high precision propagation required to simulate the low amplitude wavepacket that leaves the nucleus. Moreover, the phase space representation allows one to identify by eye the mechanisms leading to the formation of the different correlated wavepackets. Finally, {\pvb} computations are straightforward to parallelize to multi-cores and multi-machine environments. {\pvb} also requires very little initial tuning: only the desired accuracy and the phase space size have to be defined.

The main current limitation of {\pvb} lies in the size of the reduced basis. This limitation can potentially be overcome by decomposing the basis in terms of sums of product of one dimensional terms. Therefore, it is interesting to consider the possibility of merging {\pvb} with MCTDH. We note that the two methods, MCTDH and {\pvb}, use fundamentally different approaches: MCTDH aims at separating the degrees of freedom while {\pvb} focuses on reducing the effective grid size. Since the two methods do not have the same domains of efficiency, it is very possible that merging the two methods could provide a method with the advantages of both approaches, able to simulate both high dimensional and long range dynamics. For instance, one may consider replacing the one dimensional discrete variable representation of MCTDH by the {\pvb} method, or alternatively, using {\pvb} to describe each electron in 3-d and introducing the correlation via MCTDHF, i.e. MCTDH with exchange symmetry built in via the Slater determinants.

The {\pvb} code developed for this project is available upon request and we would be happy to cooperate in further developing the methodology and testing new applications.

\begin{acknowledgments}
 We wish to thank Daniel Pelaez-Ruiz, Matthieu Sala and Hans-Dieter Meyer for their
invaluable help and guidance in utilizing the MCTDH code
\cite{MCTDH-software} and optimizing its use for our purposes. This work was supported by the Israel Science Foundation ($533/12$), the Minerva Foundation with funding from the Federal German Ministry for Education and Research and the Koshland Center for Basic Research.
\end{acknowledgments}


\end{document}